\documentclass[twocolumn]{article}
\usepackage[acmtitlespace,nocopyright]{acmneww}

\usepackage{graphicx}
\usepackage{amssymb,amsfonts,amsmath}

\newcommand{\R}{\mathbb{R}}

\title{Hierarchical ordering of reticular networks}

\author{
Yuriy Mileyko\thanks{Department of Mathematics,
Duke University, Durham, NC 27708, USA},
Herbert Edelsbrunner\thanks{Institute of Science and Technology Austria, 3400 Klosterneuburg, Austria}
\thanks{Department of Computer Science,
Duke University, Durham, NC 27708, USA},
Charles A. Price\thanks{School of Plant Biology, The University of
Western Australia, Crawley WA 6009, Australia}, \and
Joshua S. Weitz\thanks{School of Biology and School of Physics,
Georgia Institute of Technology, Atlanta, GA 30332, USA}}

\begin{document}
\maketitle

\begin{abstract}
{\small
The structure of hierarchical networks in biological and
physical systems has long been characterized using the Horton-Strahler
ordering scheme.
The scheme assigns an integer order to each edge in the
network based on the topology of branching such that the order increases
from distal parts of the network (e.g., mountain streams
or capillaries) to the ``root'' of the network (e.g., the river outlet or the aorta).
However, Horton-Strahler ordering cannot be applied
to networks with loops because they 
they create a contradiction in the edge ordering
in terms of which edge precedes another in the hierarchy.
Here, we present a generalization of the Horton-Strahler order to weighted
planar reticular networks, where weights are assumed to correlate with
the importance of network edges, e.g., weights estimated from
edge widths may correlate to flow capacity. Our method assigns hierarchical levels
not only to edges of the network, but also to its loops, and
classifies the edges into reticular edges, which are responsible for
loop formation, and tree edges. In addition, we perform a detailed and
rigorous theoretical analysis of the sensitivity of the hierarchical
levels to weight perturbations. We discuss applications of this
generalized Horton-Strahler ordering
to the study of leaf venation and other biological networks.
}
\end{abstract}
\vspace{0.1in}

\section*{Introduction}
Networks and network theory have been utilized to 
represent and analyze the structure
and function of a myriad of biological systems.  These
systems span scales from cells to ecosystems and
include gene regulatory
networks~\cite{barabasi_nat2004,alon_2007}, 
metabolic pathways~\cite{guimera_nat2005,duarte_2007}, disease
dynamics~\cite{newman_pre2002,meyers_2005}, food webs~\cite{pascual_book2006,allesina_sci2008}, 
host-parasite webs~\cite{lafferty_pnas2006,flores_pnas2011}, and 
social interactions \cite{watts_nat1998,dodds_sci2003,christakis_2007}.
In the process, structural archetypes have been identified
including scale-free behavior, motifs,
modularity, the emergence of hubs, and small-world
structure~\cite{watts_nat1998,barabasi_sci1999,strogatz2001,girvan_2002,albert_rmp2002,newman_siam2003,newman_pnas2006,alon_book2007}.
However, these theories do not typically
incorporate the spatial constraints that underlie
the location and connections amongst nodes and edges.
Indeed, there are many examples of
delivery and distribution networks where nodes and edges
are physical structures embedded in space, e.g.,
leaf venation networks \cite{turcotte_jtb1998,brodribb_2010}, cardiovascular networks \cite{labarbera_1990,kassab_2006},
cortical networks \cite{zhang_pnas2000},
root networks \cite{waisel_2002},
ant trails \cite{latty_2011} and road networks \cite{masucci_2009}.
Hence, theory is also needed to characterize
biological networks whose structure is strongly influenced by physical constraints
(for a review, see \cite{barthelemy_2011}). 

Although the theory of spatial networks is quite diverse, the theory
as applied to resource delivery networks in biology often involves
certain simplifying assumptions.
For example, in fractal branching theory, 
a network is seen as a perfectly self-similar
structure, e.g. a dividing binary tree \cite{rashevsky62}. A prominent 
theory of metabolic scaling in mammals 
assumes the cardiovascular system
is a fractal whose physical dimensions
have evolved to optimally transport fluid from the aorta to capillaries \cite{west_sci1997,brown_eco2004}.
An extension of this model to
the above-ground structure of tree branches makes similar 
assumptions \cite{enquist_nat1998}.
Both models 
have inspired a wide array of follow-up work
with increased recognition that the original fractal
branching assumption is overly simplistic
\cite{dodds_2001,price_2007,savage_2008,banavar_2010,kolokotrones_nat2010}.
%
For example, in reality, physical networks in biology have 
side branches and are not perfectly balanced
binary trees \cite{turcotte_jtb1998}.  
Theories of side-branching resource delivery and distribution networks
have their origins in the study of river networks.
In a river network, streams merge together to form larger streams.
However, small streams can merge into larger streams of all scales.
The topological structure of river networks can be analyzed using
the so-called Horton-Strahler order \cite{horton1945, strahler1957}.
This scheme assigns an
integer number to every branch of the network. 
The numbers represent
different levels of the branch hierarchy, with larger numbers
corresponding to the larger stream segments in the network.
The Horton-Strahler ordering is the basis for the characterization
of the statistical properties of river networks\cite{dodds_2000},
including the finding
that river networks are fractal \cite{rodrigueziturbe_1997}.
Moreover, the side-branching
statistics first introduced by Tokunaga \cite{tokunaga_1978}
can be used to characterize universal features
of river networks and departures thereof \cite{dodds_pre2001}.

Leaf venation networks are a 
prominent example of a physical delivery and distribution
network whose structure possess numerous side branches.
The structure of leaf venation networks has
broad functional implications.
For example, leaf vein density is positively correlated
with photosynthetic rates~\cite{brodribb_2007} 
and also influences the extent
to which leaves form a hydraulic bottleneck in whole 
plants~\cite{cochard_2004,sack_2006}.
However, many leaves of higher plants (notably most leaves
of angiosperm lineages),
have reticulate venation networks, 
involving loops within loops \cite{brodribb_2010}.
It has been hypothesized that
reticulate patterns allow leaves to maintain the supply of water and nutrients 
to and from photosynthetically active chloroplasts even when flow through some 
edges in the network is lost~\cite{nardini_2001,katifori_2010,sack_pnas2008,corson_2010} 
due to mechanical damage or herbivory. 
Unfortunately, the Horton-Strahler ordering scheme developed for
the analysis of river networks is not directly applicable to reticular networks.
The reason is that loops lead to inconsistencies in the merging procedure
in which a strictly
hierarchical order is assigned to all edges.

In this paper we propose a method that generalizes the Horton-Strahler order
to planar, weighted reticular networks. Such networks encompass a large
class of physical networks, where weights can often be obtained by
estimating dimensions of edges, such as branch widths, 
or other indicators of cost or importance. 
While coinciding with the Horton-Strahler
order for branching networks, our method also assigns hierarchical levels
to the loops of the network. Moreover, it categorizes the branches into
the ones responsible for the formation of loops, and the ones forming
the tree structure of the network. Edge weights play an important role in our
algorithm, and we perform a theoretical analysis of possible effects of
weight perturbations on the hierarchical levels.  We find that
the loop hierarchy is more robust to measurement error of network edge
weights than is the tree hierarchy.  
In the past, comparisons of the statistical similarity between
river networks and leaves have been proposed, albeit such comparisons
are restricted to leaves without loops~\cite{pelettier_2000}.
Hence, we also discuss applications of
the current method to the characterization and comparison
of reticulate leaf venation networks as well as obstacles to extending 
this method to a more general class of networks.

\section*{A graph theoretic approach to Horton-Strahler ordering of rooted trees}
\begin{figure}[tb]
\begin{center}
\includegraphics[width=0.5\textwidth]{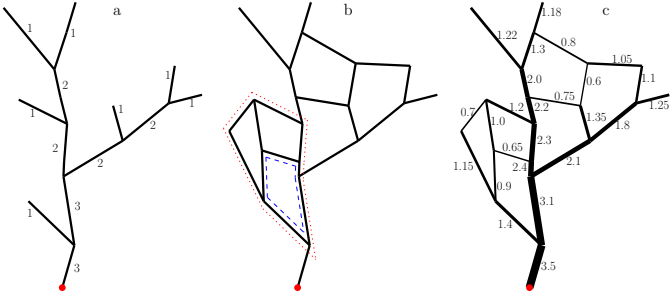}
\caption{\label{fig:strahler} \footnotesize Examples of networks with hierarchical
structure with a common ``root'' or outlet denoted by
the red dot at the bottom of each network: (a) Horton-Strahler stream
order of branch hierarchy in a tree network; (b) Reticular network with
possible loop hierarchy: the, blue, dashed loop might be less important than
the red, dotted loop; (c) Reticular network of (b) with weights.}
\end{center}
\end{figure}

We start by reviewing the algorithm for constructing the Horton-Strahler
order. For the remainder of the paper, we shall adopt the language of graph
theory \cite{chartrand_book1985, bollobas_book1998}. Note that in 
graph theory, the ``leaves'' of the network are those vertices which
only have a single edge that connects to them.
In this context, the input to the Horton-Strahler
ordering algorithm is simply a rooted
tree, $T=(V, E)$, where $V$ is the set of vertices and $E$ is the set of
edges. Given such a tree, the algorithm assigns a level, $\lambda(e)$, to
each edge $e\in E$ in the following way:
\begin{itemize}
\item Assign level $1$ to all edges connected to the leaves of $T$.
\item For each vertex having only one incident edge, $e$, with undefined
$\lambda(e)$, let $l$ be the maximal level
among the other incident edges. If there is a single incident edge
of level $l$, then $\lambda(e)=l$. If there are two or more incident
edges of level $l$, then $\lambda(e)=l+1$.
\end{itemize}
The result of this algorithm is illustrated in Fig. \ref{fig:strahler}(a).
Conventionally in the study of river networks~\cite{rodrigueziturbe_1997},
this algorithm can be summarized by a single rule which states
that the order of a downstream segment is equal to
$$
\lambda = \textrm{max}(\lambda_1,\lambda_2)+\delta_{\lambda_1,\lambda_2}
$$
where $\lambda_1$ and $\lambda_2$ are the order of 
the two upstream segments that are merging and $\delta$ is the Kronecker
delta.

It is clear, however, that if the network has loops, as in Fig.
\ref{fig:strahler}(b), the algorithm simply cannot proceed because there
always will be a vertex having more than one incident edge with an
undefined level. Moreover, loops in this graph seem to also form a
hierarchy. For example, the loop outlined in Fig. \ref{fig:strahler}(b)
by the red dotted line may belong to a higher level than the loop
outlined by the blue dashed line. It turns out that such a hierarchy
can be constructed and separated from the tree hierarchy if edges
have weights and the graph itself is planar. An example of such a
graph is shown in Fig. \ref{fig:strahler}(c), where the weights
represent widths of the branches.

To guide the reader's intuition, we first provide an alternative description
of the Horton-Strahler algorithm for the case when the tree $T$ is
binary and weighted. Let $w:E\to\R$ be the weight function, that is,
$w(e)$ is the weight of an edge $e\in E$. Since the tree is rooted, there is a partial
order defined on $E$ as follows: $e_1\leq e_2$ if there is a path from
the root to $e_1$ which passes through $e_2$ (in other words, $e_2$ is
closer to the root than $e_1$).  Let us assume that the weight function
is strictly increasing with respect to this order, that is,
$w(e_1)<w(e_2)$ if $e_1<e_2$. In this case, the Horton-Strahler order can
be computed using the following procedure:
\begin{itemize}
\item Let $C\subset E$ be the set of edges incident to leaves of $T$,
regarded as a set of disjoint components. For each $e\in C$
let $\lambda(e)=1$.
\item Iterate through (the rest of the) edges in order of increasing
weight. For each edge $e$ do the following:
\begin{itemize}
\item If $e$ shares a vertex with a single component $c_1\in C$, then
merge $c_1$ and $e$ into a new component $c$, and let
$\lambda(e)=\lambda(c)=\lambda(c_1)$.
\item If $e$ shares a vertex with two components $c_1, c_2\in C$, then
merge $c_1, c_2$, and $e$ into a new component
$c$, and assign levels as follows:
\begin{itemize}
\item If $\lambda(c_1)=\lambda(c_2)$, then $\lambda(c)=\lambda(c_1)+1$.
\item If $\lambda(c_1)\neq\lambda(c_2)$, then
$\lambda(c)=\max{\{\lambda(c_1), \lambda(c_2)\}}$.
\item $\lambda(e)=\lambda(c)$.
\end{itemize}
\end{itemize}
\end{itemize}

\section*{Ordering of planar weighted graphs}
We now present the algorithm for constructing the generalized
Horton-Strahler order.
Let $G=(V, E)$ be a planar graph, not necessarily a tree, and again
let $w:E\to \R$ be a weight function. We shall assume that $w$ is
injective (i.e., all weights are unique).  
Otherwise, the ties will be resolved arbitrarily. The
merging procedure for computing the Horton-Strahler order works with
disjoint components, which, in the language of algebraic topology, are
$0$-dimensional homology classes. Loops, on the other hand, are
$1$-dimensional homology classes. Hence, we may try to construct a
hierarchy by merging loops. Notice that the boundary of a face of the graph
$G$ is a loop, and we can merge two neighboring faces by removing a
shared edge. Using these two observations, we obtain
the following merging procedure for loops:
\begin{itemize}
\item Sort the edges so that $w(e_1)<w(e_2)<\cdots<w(e_n)$,
where $n=|E|$ is the number of edges.
\item Let $\lambda(f)=1$ for each face $f$.
\item Iterate through $e_1,\ldots, e_n$ and do the following:
\begin{itemize}
\item If $e_i$ is adjacent to a single face, skip to the next edge.
\item If $e_i$ is adjacent to two distinct faces $f_L$ and $f_R$,
remove $e_i$ from the graph, let $f_{merged}=f_L\cup f_R$, and assign
the levels as follows:
\begin{itemize}
\item If $\lambda(f_L)=\lambda(f_R)$ then
$\lambda(f_{merged})=\lambda(f_L)+1$.
\item If $\lambda(f_L)\neq\lambda(f_R)$ then
$\lambda(f_{merged})=\max\{{\lambda(f_L), \lambda(f_R)}\}$.
\item $\lambda(e_i)=\min\{{\lambda(f_L), \lambda(f_R)}\}$.
\end{itemize}
\end{itemize}
\end{itemize}

\begin{figure}[tb]
\begin{center}
\includegraphics[width=0.5\textwidth]{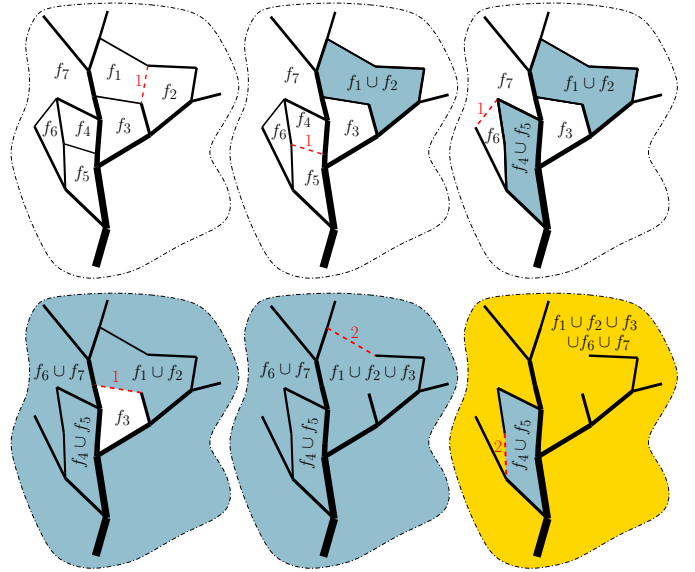}
\caption{\label{fig:merge} \footnotesize  An illustration of the loop merging procedure
applied to the graph from Fig. \ref{fig:strahler}(c). Red, dashed edges
are the ones removed during merging, the corresponding numbers show
their levels. Levels of faces is encoded by the color: white faces have
level $1$, light blue faces have level $2$, and gold faces have level $3$.
Note that $f_7$ is the unbounded face.}
\end{center}
\end{figure}

A step-by-step illustration of loop merging applied to the tree in Fig.
\ref{fig:strahler}(c) is shown in Fig.  \ref{fig:merge}. Notice that
this procedure builds a rooted binary tree, where leaves correspond to
the faces of $G$, and the rest of the vertices correspond to unions of
these faces. The assignment of levels in this tree follows the original
Horton-Strahler algorithm.  It is also useful to remember that faces of
$G$ are vertices of its dual graph, $G^*$, and merging faces of $G$ can
be thought of as adding an edge to $G^*$. Hence, the two merging
procedures that we described are, in some sense, dual.  We shall refer
to the binary tree of faces as the \emph{co-tree} of $G$, and denote it
by $T^*(G)$.

\begin{figure}[tb]
\begin{center}
\includegraphics[width=0.5\textwidth]{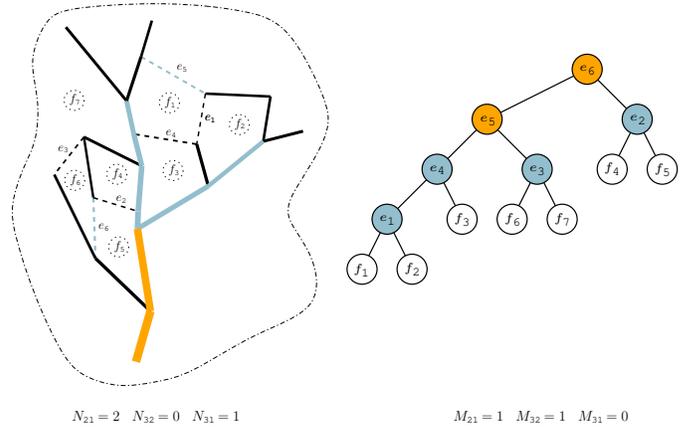}
\caption{\label{fig:merge_result}  \footnotesize Hierarchical levels assigned to the
loops and branches of the network from Fig. \ref{fig:strahler}(c). Edge
levels are shown on the left, where black edges have order $1$, light
blue edges have order $2$ and gold edges have order $3$; reticular edges
are dashed. Face levels are shown in the co-tree on the right, where
white nodes have order $1$, light blue nodes have order $2$ and gold
nodes have order $3$. Leaves of the co-tree are labeled by the
corresponding faces while other nodes are labeled by the reticular edges
causing the merger of the two child nodes. Numbers $N_{ij}$ are Tokunaga
statistics for the spanning tree and indicate the number of edges of
level $j$ joining with edges of level $i$~\cite{tokunaga_1978}.  Similarly, $M_{ij}$ are
Tokunaga statistics for the reticulate co-tree and indicate the number of 
edges and
faces of level $j$ merging with edges of level $i$. For both $M$ and $N$,
statistics are only collected when $i>j$.}
\end{center}
\end{figure}

The construction of $T^*(G)$ removes edges from $G$ which are
responsible for the existence of loops. We shall call such edges
\emph{reticular}. Assignment of levels for such edges is based on the
assumption that a merger should not be more significant than any of the
merging elements. Notice that after removing reticular edges from $G$ we
have a spanning tree of $G$, which we denote by $T(G)$. This tree
captures the tree-like structure of the original network, and we can
assign hierarchical levels to its edges using the original
Horton-Strahler algorithm. We only need to determine which vertex should
be the root, and we do this by finding the vertex with a single incident
weight of maximum weight. Hence, we augment the procedure for constructing the
loop hierarchy by the following statement:
\begin{itemize}
\item Apply the Horton-Strahler ordering to the remainder of the graph
(which is a rooted tree).
\end{itemize}
The result of the complete algorithm applied to the tree in Fig.
\ref{fig:strahler}(c) is provided in Fig. \ref{fig:merge_result}.

The algorithm produces three types of output.  First, it provides
a unique set of orders to those edges involved in the non-reticulate
component of the network (Figure 3 - left panel).  Second,
it provides a unique set of orders to those edges 
involved in the formation of loops (Figure 3 - right panel).  
Further, one can also calculate the side-branching statistics
associated with both orderings.  The side-branching statistics,
i.e., ``Tokunaga'' statistics~\cite{tokunaga_1978},
for a conventional non-loopy tree are
summarized by the numbers $N_{ij}$ which are the number of edges
of level $j$ that join with edges of level $i$.  Because of
the ordering process, these statistics are evaluated for
$i>j$.  These numbers
can also be divided by the number of absorbing edges, i.e., the total
number of edges of level $i$ to yield an average number of
side-branches per segment.  Here, the algorithm produces two
sets of Tokunaga statistics, the numbers $N_{ij}$ for the
side-branching of tree edges (Figure 3 - left panel) 
and $M_{ij}$ for the side-branching of reticulate edges (Figure 3 - right panel).

\begin{figure}[tb]
\begin{center}
\includegraphics[width=0.5\textwidth]{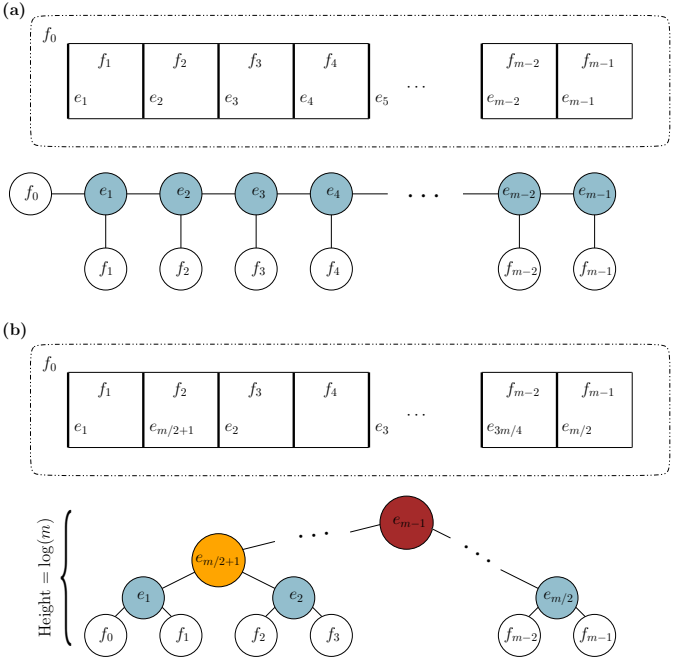}
\caption{\label{fig:worst_case}  \footnotesize Example of the two extreme cases of the
loop hierarchy. The network has $m$ faces, where $m=2^{k}$ for some
integer $k>0$. $(m-1)$ of these faces are adjacent squares and the other
one is the unbounded face.  Vertical edges are removed before horizontal
edges as follows: (a) The edges are removed sequentially from left to
right.  The corresponding co-tree has the shape of a ``comb'' and the
maximal hierarchical level is $2$; (b) The edges are removed from left
to right skipping every second edge.  The process is repeated until all
vertical edges except the rightmost one are removed. The corresponding
co-tree has the height $log(m)=k$ which is the maximal hierarchical
level.}
\end{center}
\end{figure}

\section*{Sensitivity of planar network ordering to weight perturbations}
Clearly, edge weights play an important role in the construction of both
loop and tree hierarchies. Unfortunately, weight estimation done in
practice is often imprecise, so the order in which the algorithm
iterates through the edges may be perturbed. In this section we
investigate how such a perturbation affects the loop and tree
hierarchies.

We start by considering the worst possible change in the hierarchical
levels of loops. Notice that the highest level in the hierarchy of loops
can be as low as $2$. This happens when the first reticular edge creates
a level $2$ face and every other reticular edge merges a level $1$ face
with the only level $2$ face (see Fig.  \ref{fig:worst_case}(a) ). On
the other hand, the highest level in the loop hierarchy can be as high
as $\log(m)$, where $m$ is the number of faces. This happens when level
$1$ faces are merged only with level $1$ faces until only faces of level
$2$ are left, then level $2$ faces are merged with level $2$ faces until
only faces of level $3$ are left, and so on (see Fig.
\ref{fig:worst_case}(b) ). It is clear from the example in Fig.
\ref{fig:worst_case} that there is a permutation of edges that can change
the loop hierarchy from one of the extreme cases to the other. However,
in practice such a permutation would generally result in from a
significant perturbation in weights. For small perturbations, it is more
likely that only a few transpositions of edges will occur.

Let $e_1,\ldots,e_n$ be the order of edges with respect to their
weights.  We shall now analyze how the structure of $T(G)$ and $T^*(G)$
changes when a single transposition occurs, that is, when the order of
$e_i$ and $e_{i+1}$ is swapped. First, we notice that there will be no
changes to the structure of the co-tree or the spanning tree if $e_i$
and $e_{i+1}$ are both tree edges, or if $e_i$ is a tree edge and
$e_{i+1}$ is a reticular edge.  Hence, there are two cases to consider:
when both $e_i$ and $e_{i+1}$ are reticular, and when $e_i$ is a
reticular edge and $e_{i+1}$ is a tree edge. In the former case, we can
regard reticular edges as edges of the co-tree. We see then that
swapping the two edges may shift a subtree of the co-tree only one level
up or down. Therefore, it is reasonable to expect that hierarchical
levels of loops will change at most by one. The case of a reticular edge
and a tree edge is more complicated. Such a transposition may lead to
detaching a subtree of the remaining spanning tree and attaching it at a
different place. This may have a drastic effect on the tree hierarchy. A
detailed analysis of the two cases justifying the above conclusions is
present below.

\begin{figure*}[tb]
\begin{center}
\includegraphics[width=\textwidth]{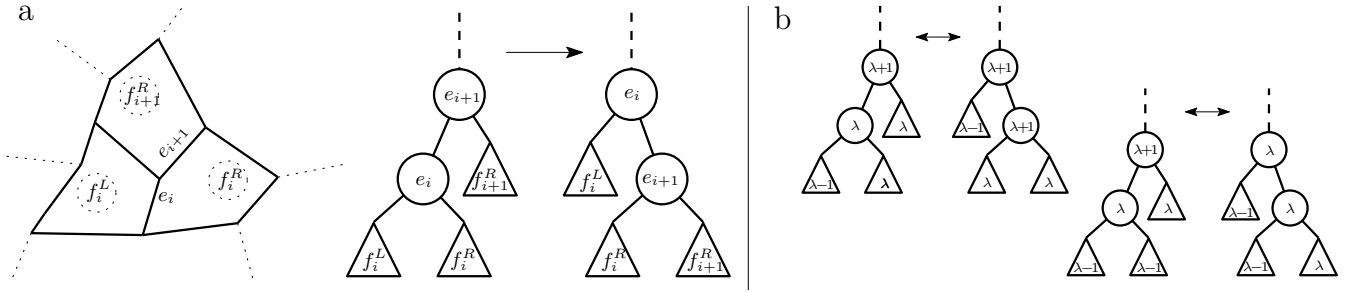}
\caption{\label{fig:case_1}  \footnotesize Example of the effect of a single
transposition of two reticular edges: (a) the part of the network
containing the two edges being transposed and the effect of the
transposition on the structure of the co-tree; (b) possible level
changes caused by the transposition.}
\end{center}
\end{figure*}

\paragraph{Case 1.} $e_i$ and $e_{i+1}$ are both reticular. Only the
co-tree can be affected in this case. Let $f_i^R$, $f_i^L$ and
$f_{i+1}^R$, $f_{i+1}^L$ be the faces merged by removing $e_i$ and
$e_{i+1}$, respectively. Also, let $f_i=f_i^L\cup f_i^R$ and
$f_{i+1}=f_{i+1}^L\cup f_{i+1}^R$. Notice that if $f_i\neq f_{i+1}^L$
and $f_i\neq f_{i+1}^R$, then $f_i$ is not a child of $f_{i+1}$ in
$T^*(G)$), and there will be no changes to the structure of the co-tree.
Suppose that $f_i=f_{i+1}^L$ (the case when $f_i=f_{i+1}^R$ follows the
same argument). Then $e_{i+1}$ is adjacent to either $f_i^L$ or $f_i^R$;
let us assume it's $f_i^R$. Removing $e_{i+1}$ before $e_i$ leads to
merging $f_{i+1}^R$ with $f_i^R$ first, and then merging the resulting
face with $f_i^L$. The corresponding change in the tree structure, shown
in Fig \ref{fig:case_1}, is a single rotation around $f_{i+1}$.
Possible changes in the levels of the nodes involved in the rotation are
also shown in Fig. \ref{fig:case_1}. We can see that these levels can
change at most by one. However, in the worst case the change in levels
may propagate up $T^*(G)$ all the way to the root.

\begin{figure*}[tb]
\begin{center}
\includegraphics[width=\textwidth]{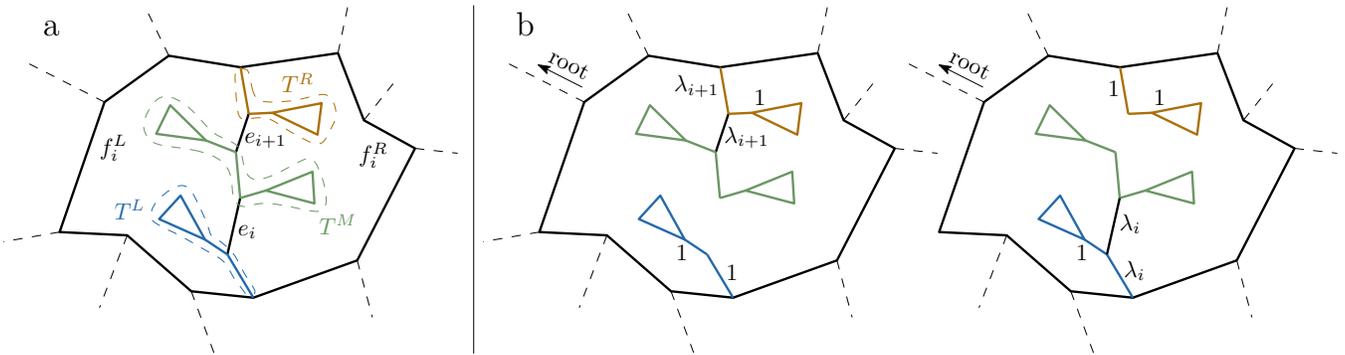}
\caption{\label{fig:case_2}  \footnotesize Example of the effect of a single
transposition of a reticular edge and a tree edge: (a) The part of the
network containing the two edges being transposed. The brown, blue, and
green triangles (and edges) denote the subtrees adjacent to the edges.
(b) The effect of the transposition on the structure of the spanning
tree and its hierarchical levels.}
\end{center}
\end{figure*}

\paragraph{Case 2.} $e_i$ is a reticular edge and $e_{i+1}$ is a tree
edge. Let $f^L_i$ and $f^R_i$ be the two faces merged by removing $e_i$.
Notice that there will be no changes in the structure of $T(G)$ or
$T^*(G)$ if $e_{i+1}$ is not adjacent to both $f^L_i$ and $f^R_i$.  So,
let $e_{i+1}$ be adjacent to $f^L_i$ and $f^R_i$. Then removing
$e_{i+1}$ before $e_i$ merges the same $f^L_i$ and $f^R_i$, so no
changes to the structure of the co-tree happen.  However, $e_{i+1}$
turns into a reticular edges, and $e_i$ becomes a tree edge.
Consequently, the structure of the spanning tree changes.  Let $E^{L,R}$
be the set of edges incident to both $f^L_i$ and $f^R_i$, and let
$T^{L,R}$ be the tree formed by the edges in $E^{L,R}$ and the edges
connected to $E^{L,R}$ and having only $f^L_i$ or $f^R_i$ as an adjacent
face (see Fig. \ref{fig:case_2}). Removing $e_{i+1}$ and $e_i$ splits
$T^{L,R}$ into three trees, $T^L$, $T^R$, and $T^M$, such that $T^L$ and
$T^R$ are connected to the boundary of $f^L_i\cup f^R_i$, and $T^M$ is
not (Fig. \ref{fig:case_2}). If $e_i$ is removed before $e_{i+1}$, then
$T^M$ is connected to $T^R$ by $e_{i+1}$. However, if the transposition
happens and $e_{i+1}$ is removed before $e_i$, then $T^M$ is connected
to $T^L$ by $e_i$ (Fig. \ref{fig:case_2}). To understand the effect of
such a change on hierarchical levels, we first assume that $T^M$ does
not contain the root of $T(G)$. Let $v^R$ be the vertex
incident to $e_{i+1}$ and $T^R$, and let $v^L$ be the vertex incident to
$e_i$ and $T^L$. Also, let $\lambda_{i+1}=\lambda(e_{i+1})$, where
$e_{i+1}$ is regarded as an edge in $T^M\cup e_{i+1}$ rooted at $v^R$,
and let $\lambda_i=\lambda(e_i)$, where $e_i$ is regarded as an edge in
$T^M\cup e_i$ rooted at $v^L$. Denote by $e^R$ the edge of $T(G)$ which
is next to $e_{i+1}$ in the path from $e_{i+1}$ to the root of $T(G)$,
and by $e^L$ the edge of $T(G)$ which is next to $e_i$ in the path from
$e_i$ to the root of $T(G)$. Then we can see that removing $e_{i+1}$
before $e_i$ can decrease the level of $e^R$ by at most
$\max\{{1,\lambda_{i+1}-1}\}$. At the same time, the level of $e^L$ can
increase by at most $\max\{{1,\lambda_i-1}\}$. In the worst case, these
changes can propagate up $T(G)$ all the way to the root. The case when
the root of $T(G)$ belongs to $T^M$ can lead to more drastic changes.
In this case, removing $e_{i+1}$ before $e_i$ leads to recomputing levels
of all edges in $T(G)-T^M$ by changing the root from $v^R$ to $v^L$.
Again, this change can then propagate further to the root of $T(G)$.

\section*{Discussion}
We have shown that the hierarchy of loops often observed in reticular
physical networks can be defined explicitly using a generalization of
the Horton-Strahler order. To obtain such a generalization we regard the
network as a weighted graph, with weights corresponding to the widths of
the network branches. Noticing that the Horton-Strahler order can be
computed by analyzing how specific disjoint components (sub-networks) of
a (non-reticular) network are merged as the edges are \emph{added} in
the order of increasing weight, we show that the hierarchical order of
loops in a weighted planar graph can then be computed by analyzing how
the \emph{faces} of the graph are merged as we \emph{remove} the edges
in the order of increasing weight. This approach naturally classifies
graph edges into reticular edges, which are responsible for loop
formation, and tree edges, which constitute a spanning tree of the
graph.  Hence, both the loop and the tree hierarchies can be computed.

Being able to compute hierarchical levels for loops creates
new possibilities for analyzing the structure of reticular networks.
By means of analogy, river networks can be compared by representing
their connectivity in terms of side-branching statistics~\cite{tokunaga_1978}.
These statistics
depict the ways in which smaller streams connect to larger
streams at all scales of the network~\cite{dodds_pre2001}.
A similar procedure could be applied to leaf networks.  For example,
the current algorithm decomposes reticulate networks
into a binary-tree for the loop hierarchy and a separate binary tree
for the tree hierarchy.  Both networks have associated
Horton-Strahler orders and therefore their structure can be
estimated using Tokunaga statistics.  Recent innovations
in software now permit semi-automated extraction of
the dimension and connectivity of entire leaf vein networks 
and the areoles that veins surround ~\cite{price_leafgui2011}.  
Hence, greater quantification of leaf vein networks 
from across a wide range of biological diversity will soon be available
for which to analyze leaf development, variation across environmental
gradients and in paleobotanical studies.
Current attempts to compare reticulate structure have largely focused 
on the density of areoles (i.e., network faces)
as a proxy for the ``loopiness'' of
the network~\cite{blonder_2011}.
The current study will provide additional metrics to compare 
the detailed branching structure of reticulate networks.

An important caveat to keep in mind when comparing reticulate network
structure is that estimating weights in physical networks is by no means a trivial
problem. Therefore, we have performed a theoretical analysis of possible
changes in the loop and tree hierarchies due to perturbations in edge
weights. We have shown that the worst possible change in the loop
hierarchy is attainable, but requires a significant perturbation of
weights. Taking into account that small perturbations are likely to cause only a
few transpositions in the order in which the edges are removed, we have
shown that a single such transposition can change the hierarchical
levels of loops at most by one. We have also shown that the change in
the hierarchical levels of the remaining spanning tree can be
arbitrarily large even when a single transposition is performed. It is
important to note that in either case the change does not happen for
every transposition. Rather, the transposed edges have to satisfy a
particular condition, which may happen rarely in practice. The
latter claim is supported by the numerous successful applications of the
Horton-Strahler order. While the method itself does not depend on any
weights, the connectivity of the network is obtained by analyzing
digital elevation map which contain noise~\cite{tarboton_1991,peckham_1995}.
In particular, the difference between the correct and the computed
connectivity may be exactly the same as the difference in the
connectivity of our spanning tree caused by transposing two
edges. Hence, the resulting hierarchy may be drastically different from
the correct one. Nevertheless, the Horton-Strahler order has been
successfully used for over five decades despite the potential
instability identified here\cite{horton1945,strahler1957,tarboton_1991,rodrigueziturbe_1997,dodds_2000}.
We suggest that empirical characterizations of reticulate
planar networks include randomization analysis on edge weights
to identify the robustness of claims regarding statistical structure
of side-branching of the tree and co-tree.

Many biological and physical systems are represented by non-planar physical 
networks ~\cite{gastner_2006,barthelemy_2011}
and computing hierarchical levels of loops in such networks is still an
open question. While our method can be applied to obtain the tree
hierarchy of such networks, the loop hierarchy cannot be computed in
this case because
the algorithm relies on the fact that any loop in a planar network corresponds to a
union of faces. In the mathematical language, (boundaries of) faces of a
planar graph form a canonical basis for loops ($1$-dimensional
homology classes). Such a canonical basis is not present in
non-planar graphs. It is not clear at this point how to handle the
non-planar case. Perhaps a method for computing loop hierarchies which
is not based on the widths of the network branches could provide an
answer. We hope that our approach of using algebraic topology language
to deal with nodes and loops of networks will prove useful in developing
such a method and complement other approaches.

\section*{Acknowledgments}
This work was supported by the National Science Foundation
Plant Genome Research Program (grant 0820624 to
H.E. and J.S.W.) and the Defense Advanced Projects Research Agency
(grant HR0011-09-1-0055 to J.S.W.).  Joshua S.~Weitz, Ph.D., 
holds a Career Award
at the Scientific Interface from the Burroughs Wellcome Fund.
During preparation of this manuscript the authors became aware
of a related work by Katifori and Magnasco, concurrently submitted
for publication.

\end{document}